\documentclass[%
 reprint,
superscriptaddress,
 amsmath,amssymb,
 aps,
]{revtex4-2}

\usepackage{graphicx}
\usepackage{dcolumn}
\usepackage{bm}
\usepackage{hyperref}
\usepackage{xcolor} 
\usepackage{amsmath}
\usepackage{relsize}
\usepackage[normalem]{ulem}

\newcommand{\x}{\mathbf{x}}
\newcommand{\F}{\mathbf{F}}
\newcommand{\s}{\mathbf{s}}
\renewcommand{\H}{\mathbf{H}}
\newcommand{\deltaeta}{\bm{\delta\eta}}
\newcommand{\deltaxi}{\bm{\delta\xi}}
\newcommand{\deltax}{\bm{\delta \x}}

\usepackage{amsthm}
\newtheorem*{proposition}{Proposition}
\begin{document}

\preprint{APS/123-QED}

\title{Dynamics-induced activity patterns of active-inactive clusters in complex networks}
\author{Anil Kumar}
\email{kumaranil@iisertvm.ac.in}
\affiliation{
School of Physics, Indian Institute of Science Education and Research, Thiruvananthapuram 695 551, Kerala, India
}%
\author{V. K. Chandrasekar}
\affiliation{Centre for Nonlinear Science \& Engineering, School of Electrical \& Electronics Engineering, SASTRA Deemed University,
Thanjavur 613401, Tamil Nadu, India
}
\author{D. V. Senthilkumar}
\email{skumar@iisertvm.ac.in}
\affiliation{
School of Physics, Indian Institute of Science Education and Research, Thiruvananthapuram 695 551, Kerala, India
}%

\date{\today}

\begin{abstract}

Synchrony patterns describe network states in which nodes of a coupled dynamical system are grouped into clusters based on synchronization between nodes. Beyond simple synchrony, synchronized clusters may also exhibit active or inactive states, and the collection of all such clusters constitutes an activity pattern. Although these patterns may arise naturally in networks with permutation symmetries, the requirement of symmetries imposes a restrictive and often unrealistic assumption, as many real-world networks lack such symmetries. In this work, we present synchrony patterns of coexisting active-inactive clusters that cannot be identified through symmetries. Considering dynamical systems in which intrinsic dynamics and coupling functions are odd functions in phase space, we identify all possible patterns a network can exhibit through symmetry breaking of identically synchronized clusters. The symmetry breaking of invariant clusters generates antisynchronized clusters, allowing active-inactive clusters to coexist. We show that while active clusters are external equitable partitions, inactive clusters can be purely dynamics-induced. Starting with a symmetry-broken state, we show that the existence of different invariant patterns is a function of coupling strength and intercluster weights. Finally, by combining synchronization manifolds with the Laplacian eigenvectors, we identify transversal perturbations for these patterns and present a stability analysis.

\end{abstract}

\maketitle

\section{Introduction} 

Synchronization plays a fundamental role in a wide range of natural and engineered systems, including biological rhythms \cite{winfree2001thegeometry}, chemical oscillations \cite{epstein1998anintroduction} and neural dynamics \cite{izhikevich2007dynamicalsystems, breakspear2017dynamicmodels}, ecological interactions \cite{hastings1997population_biology, murray2002mathematical_biology}, electrical power grids \cite{rohden2012self_organized, motter2013spontaneous_synchrony}, and collective behavior in social networks \cite{arenas2008, Pikovsky2001synchronization, BOCCALETTI2006complexnet}. Coupled dynamical systems exhibit a variety of collective states, such as chimera states~\cite{Abrams2004chimeras, bollt2023fractal, parastesh2021chimeras, Majhi2019chimera}, cluster synchronization~\cite{pecora2014cluster,schaub2016graph, sorrentino2016complete, cho2017stablechimera, lodi2021one, della2020symmetries, Zhang2020symmetry,kumar2024symmetry}, or complete synchronization~\cite{pecora1998master}, each of which constitutes a synchrony pattern, defined as a partition (or coloring) of nodes into synchronized clusters. The master stability framework has been instrumental in analyzing the stability of complete synchronization ~\cite{BOCCALETTI2006complexnet, arenas2008} and has been extended to cluster or chimera synchronization over the past few years \cite{pecora2014cluster, schaub2016graph, sorrentino2016complete, cho2017stablechimera, lodi2021one, della2020symmetries, Zhang2020symmetry}. For networks of identical oscillators, it has been shown that the invariance of these clusters is determined by underlying permutation symmetries of the network structure~\cite{pecora2014cluster}. Nevertheless, dynamically valid synchrony patterns can also emerge from external equitable partitions (EEPs) \cite{schaub2016graph} or balanced colorings of nodes \cite{Stewart2003symmetry}, even in the absence of explicit symmetries.

Clusters of synchronized nodes may exhibit \emph{active} or \emph{inactive} states, where inactive nodes remain at zero velocity while active ones display oscillatory motion. Similar to synchrony patterns, the collection of all such clusters defines an \emph{activity pattern}. Activity patterns differ fundamentally from purely synchrony-based patterns, as the activeness of nodes becomes an additional determinant of the overall pattern. In the past, the inactive state of nodes has been referred to as amplitude or oscillation death \cite{poel2015partial, zakharova2014chimera, ZOU2021quenching}. A network may therefore display partial death states—where only a fraction of clusters experience amplitude or oscillation death~\cite{KOSESKA2013oscillation, saxena2012amplitude, ZOU2021quenching}—or transition entirely into a complete death state at sufficiently strong coupling. Evidence of region-dependent activity levels has also been observed in the human brain~\cite{RATTENBORG2000behavioral, Gerdes2010brainactivations, Dedreu2019brainactivity, Chialvo2010emergent}, highlighting the potential relevance of activity-based patterns for understanding neurobiological dynamics. 

While complete amplitude or oscillation death can often be induced easily \cite{KOSESKA2013oscillation, saxena2012amplitude, ZOU2021quenching}, achieving coexistence between active and inactive clusters is substantially more challenging and remains less explored since it requires precise cancellation of fluctuations to maintain the invariance of inactive clusters. Furthermore, in contrast to purely synchrony-based patterns or complete death states, the stability of partial death states must simultaneously account for the stability of synchronization, inactive state of clusters, and antisynchronization between oscillating clusters, with antisynchronization being essential for maintaining the stability of inactive clusters~\cite{kumar2025symmetry}. Recent results demonstrate that synchrony patterns based on permutation symmetries can indeed support coexisting active–inactive states~\cite{kumar2025symmetry, poel2015partial}. In these cases, symmetry-related nodes adopt antisynchronized configurations that facilitate  cancellation of fluctuations for inactive clusters. However, the requirement of symmetries for the existence of these patterns is overly restrictive and often unrealistic, as it may not be satisfied by real-world network structures. Consequently, extending these patterns to networks that lack such symmetries is of fundamental importance.

In this work, we present synchrony patterns of co-existing active-inactive clusters in networks of identical oscillators whose underlying structure lacks permutation symmetries. We choose dynamical systems in which node dynamics and coupling functions are odd functions of the phase space, and show that dynamically valid clusters of synchronized nodes coexist in active and inactive states. In particular, our work shows that active clusters are based on EEPs, while inactive ones can be purely dynamics-induced. Starting from the complete amplitude death state, we break symmetry between synchronized nodes and systematically identify all possible patterns a network can exhibit. Some of the patterns may be transient; therefore, a method to find their existence at all times will be  presented. Finally, we perform a stability analysis of these patterns and validate the numerical simulations.

The paper is organized as follows. Section \ref{sec:a general setup} introduces identical oscillators connected through diffusive interactions. Section \ref{sec:eep} introduces EEPs and discusses the invariance of synchrony patterns induced by EEPs. Section~\ref{sec:node dynamics and coupling funct} provides sufficient conditions on node dynamics and coupling functions for generating patterns of coexisting active-inactive clusters. A method based on symmetry breaking of synchronized clusters to systematically generate all possible patterns, and the existence of these patterns is also discussed. Section \ref{sec:stability analysis} presents the stability analysis of the invariant patterns. Section~\ref{sec:sl_osc} describes Stuart-Landau oscillators, which are used in numerical simulations. Section \ref{sec:num_simul} presents numerical simulations. Section \ref{sec:stablity_num_obs_patts} presents stability analysis of patterns observed in numerical simulations, and in Section \ref{sec:conclusions}, we conclude the paper. 
\section{Identical coupled oscillators}\label{sec:a general setup}
We consider a network of identical oscillators coupled via diffusive coupling. The dynamics of an $i$th oscillator is governed by
\begin{equation}\label{eq:model}
    \dot{\x}_i=\F(\x_i)-\sigma \sum_{j=1}^N [L]_{ij} \H(\x_j), ~~~~i=1,2, \dots, N,
\end{equation}
where $\x_i \in \mathbb{R}^m$ denotes the state vector of the $i$th oscillator, $\F:\mathbb{R}^m \to \mathbb{R}^m$ describes the internal dynamics, $\H:\mathbb{R}^{m} \to \mathbb{R}^m$ is the coupling function, $\sigma  \ge 0$ is the overall coupling strength, and $N$ denotes the network size. The matrix L is the Laplacian matrix with entries $[L]_{ij} =-[A]_{ij}$ if $i \ne j$ and $[L]_{ii} =\sum_{j=1}^N[A]_{ij}$, where A represents the adjacency matrix with entries $[A]_{ij}=w_{ij} > 0$ if nodes $i$ and $j$ are connected but zero otherwise. We restrict ourselves to undirected networks characterized by $[A]_{ij}=[A]_{ji}$.
\begin{figure*}
 \includegraphics[width=\linewidth]{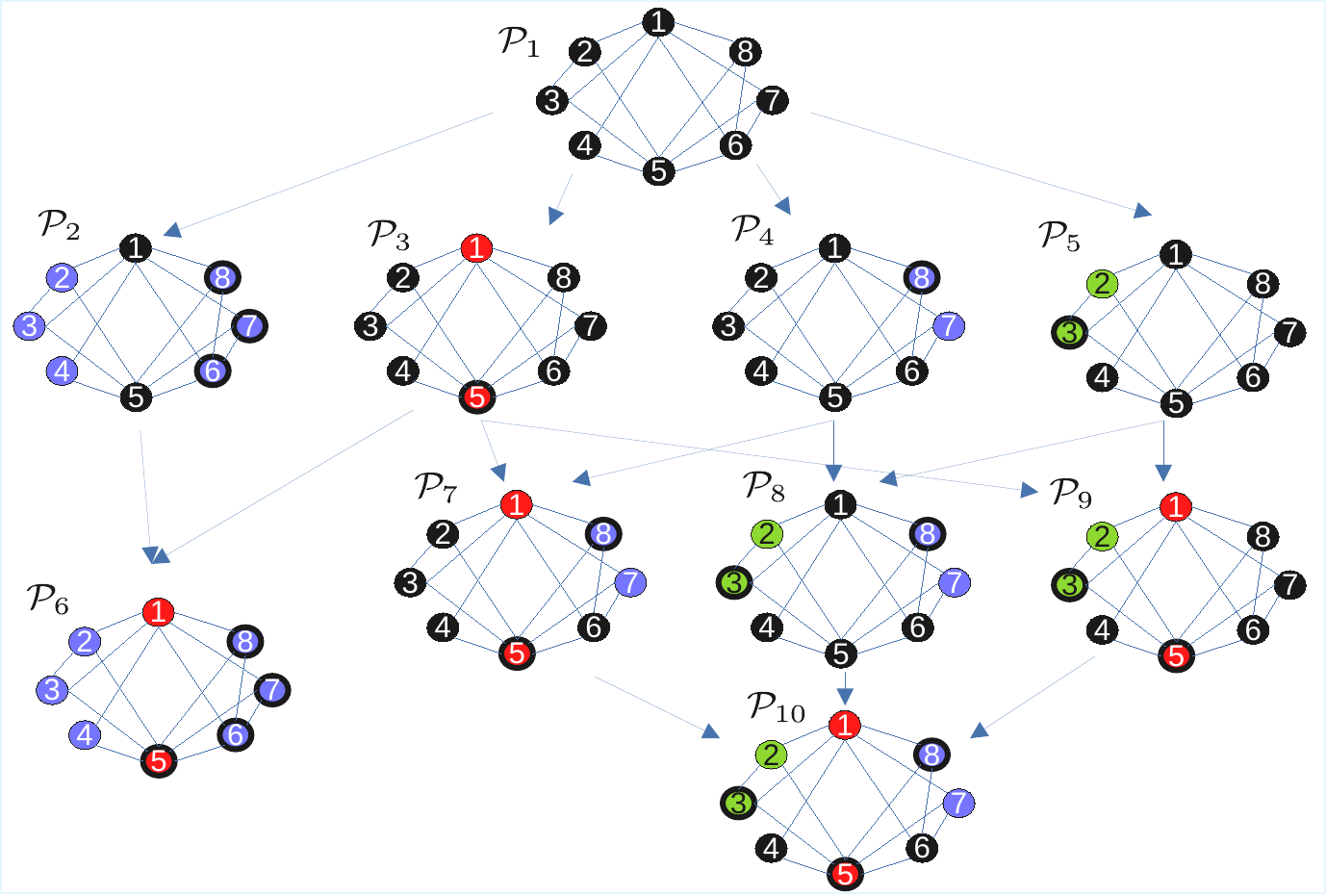}
 \caption{Coexisting active-inactive clusters are generated from symmetry breaking of identically synchronized nodes. Starting with complete amplitude death state, pattern $\mathcal{P}_1$, the symmetry breaking generates patterns $\mathcal{P}_2-\mathcal{P}_{10}$. Synchronized nodes forming a cluster are shown using the same color and boundary thickness. If two clusters are in antisynchrony with each other, one of the clusters is shown by a thicker boundary line. Inactive clusters are shown in black.}
 \label{fig:activity_patterns}
 \end{figure*}
\section{External equitable partitions}\label{sec:eep}

A network of identical oscillators described by Eq.~\eqref{eq:model} can exhibit dynamically valid
clusters of identically synchronized nodes, even when the underlying network lacks permutation symmetries. Such clusters can be identified systematically using \emph{external equitable partitions} (EEPs) \cite{schaub2016graph}, which group nodes such that every node in a cluster receives the same number of connections from each of the other clusters. Formally, suppose that an EEP partitions the network into $k$ clusters, $1 \le k \le N$, denoted by $\mathfrak{C} = \{\mathcal{C}_1, \mathcal{C}_2, \ldots, \mathcal{C}_k\}$.  
For an adjacency matrix $A$, the EEP condition requires that the nodes in any two clusters $\alpha$ and $\beta$ satisfy
\begin{align}\label{eq:eep_condition}
    \sum_{k \in \mathcal{C}_{\beta}} A_{ik}
    \;=\;
    \sum_{k \in \mathcal{C}_{\beta}} A_{jk},\;
    \forall\, i,j \in \mathcal{C}_\alpha,\; \forall\, \alpha,\beta .
\end{align}
If the initial conditions in Eq.~\eqref{eq:model} satisfy 
$\mathbf{x}_i(0)=\mathbf{x}_j(0)$ for all $i,j \in \mathcal{C}_\alpha$, identical synchrony is an invariant state for the EEP-based cluster $\mathcal{C}_\alpha$ because the intra-cluster coupling terms get canceled, and since all nodes in $\mathcal{C}_\alpha$ receive identical input from every other cluster, Eq.~\eqref{eq:model} shows that the nodes remain synchronized for all time. The extreme cases $k=1$ and $k=N$ always satisfy the EEP condition (Eq.~\eqref{eq:eep_condition}). Consequently, the fully incoherent state (perhaps transient) and the complete synchronous state, $\mathbf{x}_i(t) = \mathbf{s}(t), ~\forall~i~\in [N]:=\{1,2,\ldots,N\}$, are invariant under Eq.~\eqref{eq:model}. Each valid EEP partition generates a distinct \emph{synchrony pattern}; therefore, a single network may admit multiple synchrony patterns. 

\subsection*{Dynamics in a pattern state}

The partition of a network into EEP clusters can be represented by an indicator matrix $Q \in \mathbb{R}^{N \times k}$ whose columns encode the cluster membership of the nodes. Specifically, the \(j\)th row of the \(l\)th column is defined as $Q_{jl}=1$ if node $j \in \mathcal{C}_l$ and $Q_{jl}=0$ if node $j \notin \mathcal{C}_l$. When the system evolves on a given synchrony pattern $\mathcal{P}$, all nodes within each cluster share identical state variables, i.e., 
\begin{align*}
 \mathcal{P} = \{\x : \x_i = \s_l, ~\forall ~i \in \mathcal{C}_l, ~\forall ~l \in [k]\},
\end{align*}
where $\x=[\x_1^T, \ldots, \x_N^T]^T \in \mathbb{R}^{Nm}$ and $k$ denote the number of clusters. In this case, the full network dynamics in Eq.~\eqref{eq:model} reduces to the quotient dynamics
\begin{equation}\label{eq:quotient_dyn}
    \dot{\s}_i=\F(\s_i)-\sigma \sum_{j=1}^{k} [B]_{ij} \H(\s_j), ~~~~i=1,2, \dots, k,
\end{equation}
where $\s_i$ denotes the state vector of the $i$th cluster, and $[B]_{ij}$ represents entries of matrix $B \in \mathbb{R}^{k \times k}$ (the \emph{quotient matrix} associated with pattern $\mathcal{P}$), which describes the effective coupling from cluster $\mathcal{C}_j$ to $\mathcal{C}_i$. The matrix $B$ satisfies the relation $LQ=QB$. Applying the left inverse of Q, given by $Q^\dagger=(Q^T Q)^{-1} Q^T \in \mathbb{R}^{k \times N}$, on both sides yields
\begin{equation*}
    B = (Q^T Q)^{-1} Q^T L Q.
\end{equation*}

\section{Invariance of active-inactive states of clusters}\label{sec:node dynamics and coupling funct}
We are interested in synchrony patterns in which a fraction of clusters exhibit inactive states, i.e., $\dot \x_i = 0 \in \mathbb{R}^m, ~\forall ~i \in \mathcal{C}_l$, for at least one cluster $\mathcal{C}_l$. EEP-based partitioning of nodes into clusters only guarantees the invariance of identical synchrony between nodes in each cluster; however, for general $\F$ and $\H$, the clusters typically remain active or oscillate in time. Below, we derive sufficient conditions for a synchrony pattern to consist of active and inactive clusters. For an $i$th cluster to be inactive in a synchrony pattern $\mathcal{P}$, Eq.~\eqref{eq:quotient_dyn} requires that
\begin{equation}\label{eq:inactive_osc}
    \dot{\s}_i=0=\F(\s_i)-\sigma \{ [B]_{ii} \H(\s_i) +\sum_{\substack{j=1\\j \ne i}}^{k} [B]_{ij} \H(\s_j)\}.
\end{equation}
A sufficient condition to achieve inactivity for the $i$th cluster is to choose vector fields $\F$ and $\H$ that satisfy
\begin{equation}\label{eq:f_and_h}
\F(-\x)=-\F(\x), ~ \text{and} ~~\H(-\x)=-\H(\x). 
\end{equation}
The smoothness of the vector fields $\F$ and $\H$ ensures that $\F(0)=\H(0)=0$. If $\dot \s_i=0$, any time dependence in Eq.~\eqref{eq:inactive_osc} arises from the third term involving other clusters. A simple criterion for the state $\dot \s_i=0$ to be invariant is that $\sum_{j, j \ne i} [B]_{ij} \H(\s_j)=0$, which is possible if clusters connected with cluster $i$ split into antisynchronized states, or exhibit the fixed point state ($\x=0$), or a mixture of these states. With this configuration of neighboring clusters, Eq.~\eqref{eq:inactive_osc} admits the solution $\s_i=0$, referred to as the amplitude death state \cite{ZOU2021quenching}. More generally, the solution $\dot \s_i=0$ but $\s_i \ne 0$ may also exist, also called the oscillation death state \cite{ZOU2021quenching}. Using analogous reasoning, one could establish the invariance of antisynchronization between clusters $\mathcal{C}_l$ and $\mathcal{C}_m$. For instance, if the cluster $\mathcal{C}_l$ is connected with $\mathcal{C}_p$ and $\mathcal{C}_m$ with $\mathcal{C}_q$, antisynchrony is an invariant state for the pair if $\s_p=-\s_q$ or $\s_p=\s_q=0$. Therefore, we focus ourselves to dynamical systems whose vector fields are odd functions in the phase space. Note that these patterns must also satisfy the condition $\sum_{i=1}^N \dot \x_i=0$, failing which the coexistence of active and inactive clusters cannot be balanced. Also note that complete amplitude death, i.e., $\dot \x_i=0 ~\forall ~i \in [N]$, is always an invariant state of the dynamical systems defined by Eqs.~\eqref{eq:model} and \eqref{eq:f_and_h}.

\subsection{Symmetry breaking of invariant clusters and pattern generation}\label{sec:symmetry_breaking}
We construct activity patterns of coexisting active-inactive clusters as follows. The complete amplitude-death state is the trivial EEP partition. Starting with this state, we break identical synchrony between nodes and create antisynchronized clusters such that clusters connected with these antisynchronized clusters can remain inactive. The process is repeated until there are no symmetries in the network.  

For example, Fig.~\ref{fig:activity_patterns} displays invariant patterns of coexisting active-inactive clusters in a network of $8$ nodes. Starting with pattern $\mathcal{P}_1$, the symmetry breaking of the nodes generates patterns $\mathcal{P}_2-\mathcal{P}_{10}$. While patterns $\mathcal{P}_1-\mathcal{P}_3$, $\mathcal{P}_6$, and $\mathcal{P}_9$ are valid EEPs, inactive clusters in the rest of the patterns are purely dynamics-induced clusters; therefore, these patterns fall outside the EEP framework. However, note that active clusters in any such pattern still satisfy the EEP condition, without which their invariance cannot hold, and only the clusters in the amplitude death state can be purely dynamics-induced. 
Clusters shown in black in Fig.~\ref{fig:activity_patterns} correspond to the amplitude death state, while those in blue and red can be in either active or inactive (oscillation death) states.
Except for $\mathcal{P}_1$, patterns corresponding to the partial death states may be transient because antisynchronized clusters in the oscillatory state may manifest as  an amplitude or oscillation death state as time progresses. In the following, we provide a method to find the existence of these patterns for all time $t>0$. 

\subsection{Existence of invariant states}\label{sec:exist_of_inv_states}

As discussed above, the symmetry breaking of EEP clusters generates activity patterns consisting of partial amplitude or oscillation death states, or all clusters in oscillatory states but antisynchronized with one another. To find the $\sigma$ range in which a pattern exists for all times, we proceed in reverse order, i.e., we start with a symmetry-broken state $\mathcal{P}_i$, and the existence of this, as well as all of the preceding patterns from which $\mathcal{P}_i$ can be reached, is determined by tracking the time evolution of active clusters in $\mathcal{P}_i$. In pattern $\mathcal{P}_i$, if one or more antisynchronized pairs of clusters cannot sustain their oscillations at a given $\sigma$ and transition to an inactive state (corresponding to pattern $\mathcal{P}_j$), the pattern $\mathcal{P}_j$ exists at coupling $\sigma$.  For instance, starting with patterns $\mathcal{P}_6$ or $\mathcal{P}_{10}$ in Fig.~\ref{fig:activity_patterns}, the pattern at the end of the symmetry breaking chains, the existence of all patterns can be determined by tracking the active clusters from these states. For a pair of antisynchronized clusters $\mathcal{C}_i$ and $\mathcal{C}_j$, the cluster $\mathcal{C}_i$ satisfies 
\begin{align}\label{eq:antisynch_clust}
    \dot \x_i&=\F(\x_i)-\sigma \{B_{ii} \H(\x_i)+B_{ij}\H(\x_j)+\sum_{\substack{l=1\\ l \ne i,j}}^k B_{il} \H(\x_l)\},
\end{align}
since $\x_j=-\x_i$ and $\H(-\x_i)=-\H(\x_i)$ , the first two coupling terms depend only on $\x_i$. Now, to simplify the numerical computations, we assume that each active cluster in $\mathcal{P}_i$ receives net zero input from clusters for which $l \ne i,j$, so the third coupling term in Eq.~\eqref{eq:antisynch_clust} becomes zero. If the assumption does not hold in a pattern,  the cluster $\mathcal{C}_i$ is coupled with other clusters, and all such coupled clusters must be considered. With this assumption, if cluster $\mathcal{C}_i$ transitions to the inactive state $\dot \x_i=0$, Eq.~\eqref{eq:antisynch_clust} reduces to  
\begin{align}\label{eq:antisynch_clust_2}
    \dot \x_i&=0=\F(\x_i)-\sigma \{B_{ii} - B_{ij}\}\H(\x_i),
\end{align}
where $B_{ii}=-\{\sum_{l=1, l \ne i,j}^k B_{il}\} -B_{ij}$. If we denote $W_i=-\{\sum_{l=1, l \ne i,j}^k B_{il}\} -2B_{ij}$, Eq.~\eqref{eq:antisynch_clust_2} becomes
\begin{align}\label{eq:antisynch_clus_3}
    \dot \x_i&=0=\F(\x_i)-\sigma W_i \H(\x_i).
\end{align}
The stability of the solution in Eq.~\eqref{eq:antisynch_clus_3} at a given $\sigma$ determines if oscillations can be sustained in clusters $\mathcal{C}_i$ and $\mathcal{C}_j$. The collective state of all clusters then determines the resulting pattern in the network, which exists for all $t>0$. 

\section{Stability analysis}\label{sec:stability analysis}
The existence of a pattern $\mathcal{P}$ does not guarantee that the trajectories initialized away from $\mathcal{P}$ will converge to $\mathcal{P}$ as time increases. To determine whether nearby trajectories converge back to the pattern, we analyze the local stability of $\mathcal{P}$ through its variational dynamics. We perturb $\x \in \mathcal{P}$ so that $\x \rightarrow \x +\deltax$, where $\deltax =[\deltax_1^T, \ldots, \deltax_N^T]^T\in \mathbb{R}^{Nm}$. Linearizing Eq.~\eqref{eq:model} about $\x$ using a first-order Taylor expansion, the evolution equation of the perturbation vector $\deltax$ can be obtained as
\begin{align}\label{eq:perturb_dyn}
    \dot{\deltax}&=\biggl\{ \sum_{l=1}^{k} E^l \otimes D\F(\s_l)-\sigma \sum_{l=1}^{k} L E^l \otimes D\H(\s_l) \biggl\} \deltax, 
\end{align}
where
\begin{align*}
    D\F(\s_l)=\left. \frac{\partial \F(\x)}{\partial \x} \right \vert_{\x=\s_l}, ~~ D\H(\s_l)=\left. \frac{\partial \H(\x)}{\partial \x} \right \vert_{\x=\s_l},\\
\end{align*}
and $\otimes$ denotes the Kronecker product. The diagonal matrices $E^l, ~l \in [k]$, in Eq.~\eqref{eq:perturb_dyn} are defined such that
\begin{align*}
[E^l]_{ii}=\begin{cases}
    1 & \text{if} ~~~~ \text{i} \in \mathcal{C}_l,\\
    0 & \text{otherwise}.
\end{cases}
\end{align*}
First, we determine the stability criteria for identical synchronization between nodes in clusters.  We find transversal perturbations or modes that break the identical synchrony between nodes. To this end, we make a coordinate transformation such that $\deltax = (T \otimes I_m) \deltaeta$, where $I_m$ is an identity matrix of size $m \times m$. The matrix $T$ is obtained as follows. 

The partition of a network into $k$ EEP clusters implies that the cluster synchronization subspace $\mathcal{S}$ and its orthogonal complement $\mathcal{S}_{\perp}$ are invariant under the operator $\mathcal{L}$, i.e., $\mathcal{LS} \subseteq S$ and $\mathcal{L} \mathcal{S}_{\perp} \subseteq \mathcal{S}_{\perp}$ (Appendix~\ref{sec:app_invariant_decomp}), where $\mathcal{L} \in \mathbb{R}^{Nm \times Nm}$ represents the bracket term in Eq.~\eqref{eq:perturb_dyn}. The invariance of $\mathcal{S}$ and $\mathcal{S}_{\perp}$ further implies that, if we choose our $T$ matrix such that its first $k$ columns form a basis in $\mathcal{S}$ and the last $N-k$ columns in $\mathcal{S}_{\perp}$, the matrix $\mathcal{L}$ can be block diagonalized such that perturbations longitudinal and transverse to the subspace $\mathcal{S}$ evolve independently from each other. We select columns of $Q$ as the basis of the subspace $\mathcal{S}$, while for the basis of the transversal subspace $\mathcal{S}_{\perp}$, we select Laplacian eigenvectors. For an EEP partition, the eigenvectors of $\mathcal{L}$ can always be arranged as $V=[V_{||} ~V_{\perp}]$, where $V_{||} \in \mathbb{R}^{N \times k}$ denotes eigenvectors forming the basis in the synchronization subspace $\mathcal{S}$ and $V_{\perp} \in \mathbb{R}^{N \times N-k}$ forms the basis in the subspace $\mathcal{S}_{\perp}$. We select the part $V_{\perp}$ and combine it with $\tilde Q$ where $\tilde Q =QM$ and $M=\operatorname{diag}(1/\sqrt{|\mathcal{C}_1|}, \ldots, 1/\sqrt{|\mathcal{C}_k|}) \in \mathbb{R}^{k \times k}$ to create the matrix $T$, i.e., $T=[\tilde Q ~V_{\perp}]$. While $V_{||}$ and $\tilde Q$ both are valid bases of $\mathcal{S}$ and any of them can be used to examine the stability of identical synchrony, we are also interested in the stability of inactive states and antisynchronization between clusters, and therefore selected $\tilde Q$. Note that there exists more than one basis of $L$ if its eigenvalues are repeated, while constructing $V$, we ensure that $V_{||}$ and $V_{\perp}$ span $\mathcal{S}$ and $\mathcal{S}_{\perp}$, respectively. The columns of $V_{\perp}$ generate modes that drive the dynamics (Eq.~\eqref{eq:perturb_dyn}) out of subspace $\mathcal{S}$ and must decay for the identical synchronization to be stable. In $\deltaeta$ coordinates, Eq.~\eqref{eq:perturb_dyn} can be expressed as
\begin{align}\label{eq:perturb_dyn_eta_coordinates}
    {\dot{\deltaeta}}&=\biggl\{ \sum_{l=1}^{k} T^{-1}E^lT \otimes D\F(\s_l) \nonumber \\
    &- \sigma \sum_{l=1}^{k}  T^{-1}L E^l T \otimes D\H(\s_l)  \biggl\} \deltaeta.
\end{align}
The matrices $T^{-1}E^lT$ are block diagonals because
\begin{align*}\label{}
T^{-1}E^{l}T=
\begin{bmatrix}
\tilde Q^TE^l\tilde Q & \tilde Q^TE^lV_{\perp}\\
    V_{\perp}^TE^l\tilde Q & V_{\perp}^TE^lV_{\perp}
\end{bmatrix}.
\end{align*}
Since $\tilde Q \in S$ and $V_{\perp} \in S_{\perp}$ and are orthogonal to each other, we have $\tilde Q^TV_{\perp}=0 \in \mathbb{R}^{k \times (N-k)}$ and $V_{\perp}^T\tilde Q=0 \in \mathbb{R}^{(N-k) \times k}$, which implies that 
\begin{equation*}
\begin{alignedat}{2}
\tilde Q^TE^lV_{\perp}&{}={}\tilde Q^TV_{\perp}V_{\perp}^TE^lV_{\perp}&{}={}0 \in \mathbb{R}^{k \times (N-k)},\\
V_{\perp}^TE^l\tilde Q&{}={}V_{\perp}^TE^lV_{\perp}V_{\perp}^T \tilde Q&{}={}0 \in \mathbb{R}^{(N-k) \times k}.
\end{alignedat}
\end{equation*}
Therefore, off-diagonal terms in $T^{-1}E^lT$ are zero. Moreover, the part $\tilde Q^TE^l\tilde Q$ is in fact diagonal, i.e., $\tilde Q^TE^l\tilde Q=\operatorname{diag}(d_1, \ldots, d_i, \ldots, d_k) \in \mathbb{R}^{k \times k}$, where $d_i=1$ if $i=l$ and $d_i=0$ otherwise. Using similar arguments, we can find that the matrices $T^{-1}LE^lT=T^{-1}LT T^{-1}E^lT$ are also in block form if $T^{-1}LT$ is block diagonal, which can be proven as follows
\begin{align*}\label{}
T^{-1}LT=
\begin{bmatrix}
    \tilde Q^TL\tilde Q & \tilde Q^TLV_{\perp}\\
    V_{\perp}^TL \tilde Q & V_{\perp}^TLV_{\perp}
\end{bmatrix}.
\end{align*}
The off-diagonal terms are again zero, i.e.,
\begin{equation*}
\begin{alignedat}{2}
\tilde Q^T L V_{\perp} &{}={} \tilde Q^T V_{\perp} V_{\perp}^T L V_{\perp} &{}={} 0 \in \mathbb{R}^{k \times (N-k)}, \\
V_{\perp}^T L \tilde Q &{}={} V_{\perp}^T L V_{\perp} V_{\perp}^T \tilde Q &{}={} 0 \in \mathbb{R}^{(N-k) \times k}.
\end{alignedat}
\end{equation*}
Therefore, $T^{-1}LT$ is also in block form. If all transverse modes $\deltaeta_{k+1}, \ldots \deltaeta_N$ decay with time as $t \rightarrow \infty$, the identical synchrony between nodes in the $k$ clusters is stable. While this method may not decouple transversal modes optimally, as witnessed in methods such as irreducible representation of symmetry groups \cite{pecora2014cluster}, cluster based coordinates \cite{cho2017stablechimera}, or simultaneous block diagonalization of coupling matrices \cite{Zhang2020symmetry}, it completely decouples the perturbations longitudinal and transversal to the synchronization manifold, which is sufficient for us to determine the stability of identical synchronization in any EEP-induced synchrony pattern $\mathcal{P}$.

\subsection{Stability of inactive clusters and antisynchronization}\label{subsec:stability inactive clusters}
The modes $\deltaeta_i, ~i \in \{k+1, \ldots,N\}$, only determine the stability of identical synchrony between nodes in EEP clusters. To determine whether an inactive state for a cluster or antisynchronization between two oscillating clusters is stable, the modes transverse to these states must also decay. 
The decay of modes transverse to the synchronization subspace $\mathcal{S}$ shows that all $\deltax_i$ associated with a cluster $\mathcal{C}_k$ decay with time; therefore, $\deltax_i\rightarrow \deltax_k, ~\forall~i \in \mathcal{C}_k$ as $t \rightarrow \infty$. If the cluster $\mathcal{C}_k$ is in an inactive state (amplitude or oscillation death state), we further require that $\deltaeta_k \propto \deltax_k \rightarrow 0$ as $t \rightarrow \infty$. Next, to determine if two clusters $\mathcal{C}_l$ and $\mathcal{C}_m$ are in antisynchrony, we require $\deltaeta_l + \deltaeta_m \propto \deltax_l +\deltax_m\rightarrow 0$ as $t \rightarrow \infty$. Now, we decouple these modes, $\deltaeta_k$ and $\deltaeta_l + \deltaeta_m$, from the tangent mode $\deltaeta_l - \deltaeta_m$ for antisynchrony. 
If we write $\deltaeta=[\deltaeta_{||}^T ~\deltaeta_{\perp}^T]^T$, the first $k$ modes in $\deltaeta$ (Eq.~\eqref{eq:perturb_dyn_eta_coordinates}) evolve as follows:
\begin{align}\label{eq:perturb_dyn_longit_modes}
    {\dot{\deltaeta_{||}}}&=\biggl\{ \sum_{l=1}^{k} \tilde Q^TE^l\tilde Q \otimes D\F(\s_l) \nonumber \\
 &-\sigma  \sum_{l=1}^{k} \tilde Q^{T}LE^l \tilde Q \otimes D\H(\s_l) \biggl\} \deltaeta_{||}.
\end{align}
As shown above, the matrices $\tilde Q^TE^l\tilde Q$ are diagonal, so the first term in Eq,~\eqref{eq:perturb_dyn_longit_modes} is simply $\operatorname{diag}(D\F(\s_1), \ldots, D\F(\s_k)) \in \mathbb{R}^{km \times km}$. We can rewrite the coupling term $\tilde Q^{T}LE^l \tilde Q$ in Eq.~\eqref{eq:perturb_dyn_longit_modes} in terms of the quotient matrix $B$. Since 
\[\tilde Q^{T}LE^l \tilde Q=\tilde Q^{T}L\tilde Q\tilde Q^TE^l \tilde Q,\] 
using the relations $LQ=QB$ and $\tilde Q =QM$, we can write the part $\tilde Q^{T}L\tilde Q$ as
\[\tilde Q^{T}L \tilde Q=M^TQ^TQBM=M^{-1}BM.\] 
For an inactive cluster $\mathcal{C}_k$ and antisynchronized clusters $\mathcal{C}_l$ and $\mathcal{C}_m$, the matrix $B$ is such that $B_{kl}=B_{km}$, $B_{lk}=B_{mk}$, $B_{lm}=B_{ml}$, and $B_{ll}=B_{mm}$. Furthermore, we have 
\begin{align*}
D\F(-\x)=D\F(\x) ~~\text{and} ~~D\H(-\x)=D\H(\x). 
\end{align*}
\begin{figure*}
 \includegraphics[width=\textwidth]{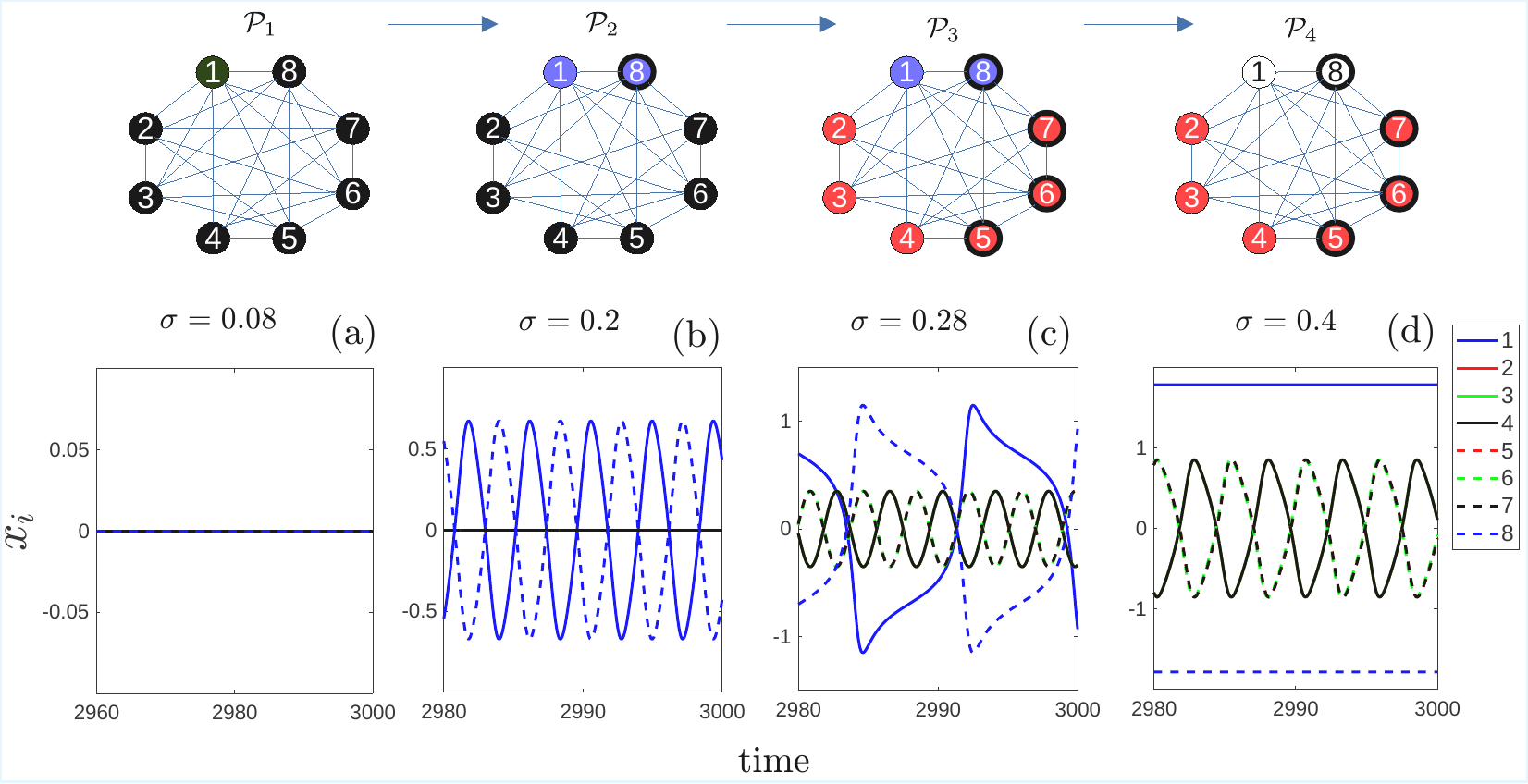}
 \caption{Time series shows activity patterns in an 8-node network of Stuart-Landau oscillators. (a)-(d) The network transitions through the patterns $\mathcal{P}_{1} \rightarrow \mathcal{P}_2 \rightarrow \mathcal{P}_3 \rightarrow \mathcal{P}_4$ as $\sigma$ increases. The parameters are $\lambda=-1, ~\omega=2$, $\sigma_x=-1$, $\sigma_y=0$. The links weights are $w_{18}=w_{81}=4$, while $w_{ij}=1$ for the rest of the links.}
 \label{fig:time_series}
 \end{figure*}
%
Because of these structural and dynamical symmetries, the tangent mode $\deltaeta_l - \deltaeta_m$ for antisynchrony evolves independently of the transversal modes. To decouple this mode, we make an additional transformation $\deltaeta \rightarrow(S^{\dagger} \otimes I_m) \deltaxi$, where  $\deltaxi =[\deltaxi_1^T, \ldots, \deltaxi_{N-n_a}^T]^T\in \mathbb{R}^{(N-n_a)m}$. The matrix $S^{\dagger} \in \mathbb{R}^{N \times (N-n_a)}$ is the right inverse of a matrix $S \in \mathbb{R}^{(N-n_a) \times N}$, where $n_a$ denotes the number of antisynchronized cluster pairs. The matrix S is derived from the identity matrix $I_{N}$, such that if $\deltaeta_l$ and $\deltaeta_m$ are summed, its $l$th row is given by $[S]_{lj}=[I_N]_{lj}+[I_N]_{mj}$, where $j \in [N]$. All other rows of $S$ are the same as those of $I_N$ except that the $m$th row of $I_N$ should be deleted. The row vectors in $S$ are then normalized to unity 2-norm. Note that $l$ and $m$ are interchangeable. The row vectors of $S$ are orthonormal, so there exists a right inverse $S^{\dagger}=S^T(SS^T)^{-1}$ such that $SS^{\dagger} =I_{(N-n_a)}$. With this transformation, Eq.~\eqref{eq:perturb_dyn_eta_coordinates} changes to 
\begin{align}\label{eq:perturb_dyn_xi_coordinates}
    {\dot{\deltaxi}}&=\biggl\{ \sum_{l=1}^{k} (T^{-1}E^lT)' \otimes D\F(\s_l) \nonumber \\
 &-\sigma  \sum_{l=1}^{k} (T^{-1}L E^lT)' \otimes D\H(\s_l) \biggl\} \deltaxi,
\end{align}
where $(.)'=S(.)S^{\dagger}$. From Eqs.~\eqref{eq:quotient_dyn} and \eqref{eq:perturb_dyn_xi_coordinates}, we calculate $(N-n_a)m$ transversal Lyapunov exponents associated with the pattern $\mathcal{P}$, and the largest of all, which we call $\Gamma$ is plotted as a function of $\sigma$. A negative value of $\Gamma(\sigma)$ shows that the pattern $\mathcal{P}$ is stable. If a network is in the complete amplitude or oscillation death state, all perturbations $\deltax_i$ must decay with time, and the stability of corresponding patterns can be determined from Eq.~\eqref{eq:perturb_dyn} itself. 
\subsection{Stability of patterns beyond EEPs} 
The stability analysis presented here is extendable to patterns that are purely dynamics induced, such as $\mathcal{P}_4$, $\mathcal{P}_{5}$, $\mathcal{P}_{7}$, $\mathcal{P}_{8}$, and $\mathcal{P}_{10}$ in Fig.~\ref{fig:activity_patterns}, but requires that clusters in amplitude death state, which are not EEP induced, should be partitioned further until a valid EEP is achieved. If there do not exist nontrivial inactive clusters satisfying the EEP condition (Eq.~\eqref{eq:eep_condition}), each inactive node in amplitude death should be treated as a separate cluster. For example, in pattern $\mathcal{P}_4$, nodes $\{1,2,3,4,5,6\}$ in amplitude death state form one cluster. For stability purposes, we should treat $\{1,5\}$, $\{2,3,4\}$, and $\{6\}$ as three clusters in the matrix $Q$, while the rest of the analysis remains the same.   
\section{Stuart-Landau oscillators}\label{sec:sl_osc}

We validate our theory using the well-known Stuart–Landau oscillators \cite{ZOU2021quenching}. The internal dynamics of a Stuart–Landau oscillator is given by  
\begin{equation*}
\F(\x_i)=\begin{bmatrix}
    \{\lambda -(x_i^2+y_i^2)\}x_i -\omega y_i \\
    \{\lambda -(x_i^2+y_i^2)\}y_i +\omega x_i
\end{bmatrix},
\end{equation*}
while for couplings, we select
\begin{equation*}
\H(\x_j)=\sigma' \begin{bmatrix}
     x_j\\
     y_j
\end{bmatrix}.
\end{equation*}
The parameter $\lambda$ determines if an oscillator is in a fixed point state ($\lambda \le 0$) or in limit cycle oscillations ($\lambda > 0$), $\omega$ determines the natural frequency of rotations, and $\sigma'=\operatorname{diag}(\sigma_x, \sigma_y) \in \mathbb{R}^{2 \times 2}$ allows for different weights for $x$ and $y$ components. Eq.~\eqref{eq:model} is solved numerically using the Runge-Kutta fourth-order method with adaptive step sizes.
\section{Numerical simulations}\label{sec:num_simul}
We present numerical simulations for the Stuart-Landau oscillators and demonstrate that dynamically valid clusters of synchronized nodes, which cannot be identified through permutation symmetries, coexist in active and inactive states. We choose a small network of eight nodes, shown in Fig.~\ref{fig:time_series}, in which the link connecting nodes $1$ and $8$ has a higher weight than other links. The heterogeneity in weights gives rise to partial death states in the network and is discussed in detail in Sec.~\ref{sec:existance_sl}.
We choose initial conditions close to pattern $\mathcal{P}_3$ and the time evolution of $x_i$ is shown in Fig.~\ref{fig:time_series}. As shown in Fig.~\ref{fig:time_series}(a), for small $\sigma$ values ($\sigma=0.08$), the network is initially in pattern $\mathcal{P}_1$, the complete death state, in which all nodes are inactive. Increasing $\sigma$ to $0.2$, the complete death state destabilizes and breaks into an inactive cluster and two antisynchronized clusters (Fig.~\ref{fig:time_series}(b)). A further increase in $\sigma$ to $0.28$ and $0.4$, the network transitions to patterns $\mathcal{P}_3$ and $\mathcal{P}_4$, respectively (Figs.~\ref{fig:time_series}(c), \ref{fig:time_series}(d)). Note that nodes $1$ and $8$ in $\mathcal{P}_4$ are in oscillation death state. All the patterns $\mathcal{P}_1-\mathcal{P}_4$ observed in Fig.~\ref{fig:time_series} are valid EEPs and are generated from the symmetry breaking of identically synchronized nodes. Although not shown in Fig.~\ref{fig:time_series}, for sufficiently large $\sigma$, the network ultimately converges to a complete oscillation death state. The partial death states (patterns $\mathcal{P}_2$ or $\mathcal{P}_4$) arise along the transition from the fully active state to complete amplitude or oscillation states. 

\begin{figure}
 \includegraphics[height=7cm, width=\linewidth]{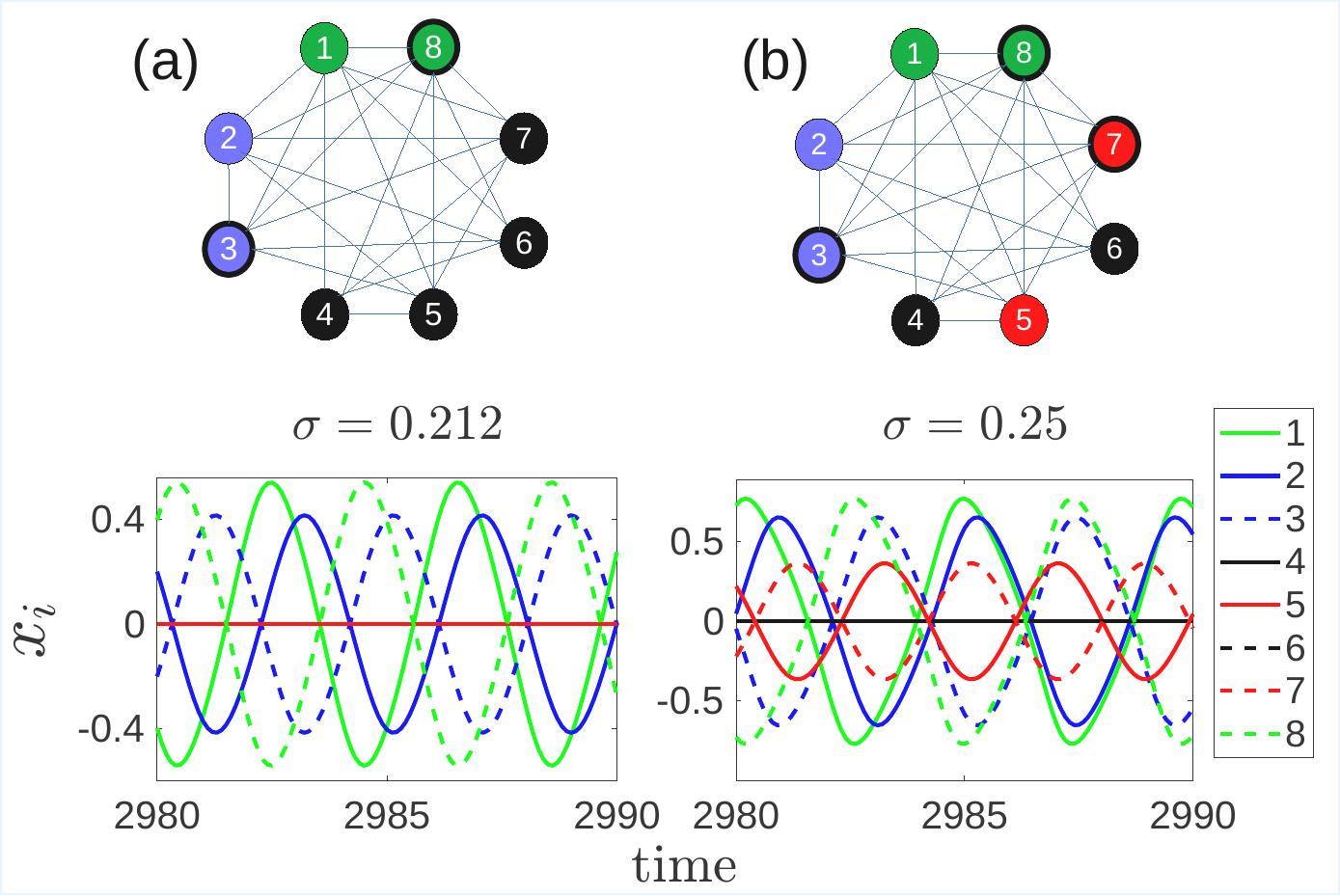}
 \caption{Time series illustrates dynamics induced clusters, which cannot be identified through EEPs, in the amplitude death state. The parameters are $\lambda=-1, ~\omega=2$, $\sigma_x=-1$, $\sigma_y=0$. The links weights are $w_{18}=w_{81}=w_{23}=w_{32}=3$, and  $w_{57}=w_{75}=2$, while $w_{ij}=1$ for the rest of the links.}
 \label{fig:dynamics_ind_patterns}
 \end{figure}

Next, we demonstrate activity patterns which originate purely from system dynamics and can not be identified through EEPs (Fig.~\ref{fig:dynamics_ind_patterns}). For a node $i$, if $\sum_{j=1, j \ne i}^N L_{ij} \H(\x_j
)=0$, the dynamical system, Eqs.~\eqref{eq:model} and \eqref{eq:f_and_h}, allows it to exhibit the amplitude death state and merge with other nodes in the amplitude death state, even if the resulting cluster is not a valid EEP.
For instance, node $4$ can merge with cluster $\{5, 6, 7\}$, Fig.~\ref{fig:dynamics_ind_patterns}(a), and with node or cluster $\{6\}$ Fig.~\ref{fig:dynamics_ind_patterns}(b), even though the resulting clusters in the amplitude death state violate the EEP condition given in Eq.~\eqref{eq:eep_condition}. Therefore, patterns outside the EEP framework arise due to purely dynamics-induced clusters in the amplitude-death state.
%
\section{Stability analysis}\label{sec:stablity_num_obs_patts}
We perform the stability analysis of patterns observed in Fig.~\ref{fig:time_series} and present the complete $\sigma$ range in which these patterns are stable. The stability of any pattern $\mathcal{P}$ requires that its corresponding quotient dynamics must exist for all $t>0$. So, first, we find the $\sigma$ range in which different invariant patterns exist in the network.
\subsection{Existence of patterns}\label{sec:existance_sl}
For Stuart-Landau oscillators, Eq.~\eqref{eq:antisynch_clus_3} changes to (ignoring the subscript $i$)
\begin{align}\label{eq:sl_steady_sol}
\begin{split}
    \dot x&=0=[\lambda -(x^2+y^2)] x -\omega y - \sigma \sigma_x W x,\\
    \dot y&=0=[\lambda -(x^2+y^2)] y +\omega x- \sigma \sigma_y W y.
    \end{split}
\end{align}
Eq.~\eqref{eq:sl_steady_sol} admits the solution $[x^* ~y^*]^T=[0 ~0]^T$, while the other solution corresponding to the oscillation death state is 
\begin{equation*}
x^*= \pm \sqrt{\frac{(\lambda-\omega c_{\mp} -\sigma \sigma_x W)}{1+c_{\mp}^2}},
y^*=  x^* c_{\mp},
\end{equation*}
where
\begin{equation*}\label{}
c_{\mp} =\frac{\sigma W(\sigma_y-\sigma_x) \mp \sqrt{\{\sigma W(\sigma_y-\sigma_x)\}^2 -4 \omega^2}}{2 \omega}.
\end{equation*}
The oscillation death state exists if  $|\sigma| \ge |2\omega/W(\sigma_y-\sigma_x)|$ and $\omega \ne0$. From Eq.~\eqref{eq:sl_steady_sol}, the stability of a fixed point solution is determined by the Jacobian matrix $J$ given by
\begin{equation*}\label{}
J=\begin{bmatrix} 
\lambda-3{x^*}^2 -y{^*}^2 -\sigma \sigma_x W & -2x^*y^*-\omega\\
-2x^*y^* + \omega & \lambda-{x^*}^2 -3{y^*}^2 -\sigma \sigma_y W
\end{bmatrix}.
\end{equation*}
Now, the eigenvalues of $J$ corresponding to $\x^*=[0~ 0]^T$ can be obtained as
\begin{equation*}\label{}
\Lambda_{1,2} =\frac{2 \lambda -\sigma W(\sigma_x+\sigma_y) \pm \sqrt{\{\sigma W(\sigma_x-\sigma_y)\}^2 -4 \omega^2\}}}{2}.
\end{equation*}
If $|\sigma|<|2\omega/(W(\sigma_x-\sigma_y))|$, $\max\!\big(\Re(\Lambda_1), \Re(\Lambda_2)\big)$ is negative if 
\begin{equation}\label{eq:amp_death_crit_coup}
2\lambda-\sigma W(\sigma_x+\sigma_y)<0,
\end{equation}
which provides the $\sigma$ range in which a cluster switches from an oscillatory state to the amplitude death state. For $|\sigma|\ge |2\omega/(W(\sigma_x-\sigma_y))|$, plotting eigenvalues $\Lambda_{1,2}(\sigma)$ as a function of $\sigma$ provides the stability insight. Similarly, the stability of the oscillation death states is determined from the eigenvalues of the Jacobian matrix $J$. The matrix $J$ reveals that the critical $\sigma$ for the transition to amplitude or oscillation death state depends on $W$; therefore, these transition points can be controlled by selecting $W$ values appropriately for different clusters. 
\subsubsection*{The example network}
We take the $8$ node network of Stuart-Landau oscillators, Fig.~\ref{fig:time_series}, and find the $\sigma$ range in which the numerically observed patterns exist. 
In the symmetry broken states $\mathcal{P}_3$, the network in Fig.~\ref{fig:time_series} exhibits EEP clusters $\mathcal{C}_1=\{1\}$, $\mathcal{C}_2=\{2,3,4\}$, $\mathcal{C}_3=\{5,6,7\}$, and $\mathcal{C}_4=\{8\}$. The clusters $\mathcal{C}_3$ and $\mathcal{C}_4$ have the same dynamics as $\mathcal{C}_2$ and $\mathcal{C}_1$, so we ignore $\mathcal{C}_3$ and $\mathcal{C}_4$ and derive the $\sigma$ range in which $\mathcal{C}_1$ and $\mathcal{C}_2$ cannot sustain their oscillations and transition to the amplitude or oscillation death state. For parameters used in Fig.~\ref{fig:time_series}, from Eq.~\eqref{eq:amp_death_crit_coup}, the amplitude death state is stable for a cluster if 
\begin{align}\label{eq:amp_death_crit_coup_2}
\sigma<\left|\frac{2\lambda}{W(\sigma_x+\sigma_y)}\right|,
\end{align}
while we find that the oscillation death state, the solution corresponding to $c_{-}$, is stable if
\begin{align}\label{eq:osc_death_crit_coup}
\sigma \ge\frac{2\omega}{W(\sigma_y-\sigma_x)}.
\end{align}
For clusters $\mathcal{C}_1$ and $\mathcal{C}_2$, $W_1=3w_{12}+3w_{13}+2w_{14}$ and $W_2=w_{21}+w_{24}+6w_{23}$, respectively, where $w_{ij}$ represents the weight of the link connecting clusters $i$ and $j$. Equations~\eqref{eq:amp_death_crit_coup_2} and \eqref{eq:osc_death_crit_coup} reveal that if $W_1 =W_2$, which is the case if $w_{ij}=1 ~\forall ~i,j$, the permissible patterns the network can exhibit are $\mathcal{P}_1 \rightarrow \mathcal{P}_3 \rightarrow \mathcal{P}_5$ as $\sigma$ increases. Partial death states arise if we consider $W_1 \ne W_2$. For instance, if $W_1 >W_2$, the permissible patterns changes to $\mathcal{P}_1 \rightarrow \mathcal{P}_2 \rightarrow \mathcal{P}_3 \rightarrow \mathcal{P}_4 \rightarrow \mathcal{P}_5$, while for $W_1 <W_2$, they are $\mathcal{P}_1 \rightarrow \mathcal{P}_6 \rightarrow \mathcal{P}_3 \rightarrow \mathcal{P}_7 \rightarrow \mathcal{P}_5$, where
\begin{align*}\label{}
\mathcal{P}_5&=\{\x_1^*, ~\x_2^*,~ \x_2^*,~ \x_2^*, ~-\x_2^*,~-\x_2^*,~-\x_2^*,~-\x_1^*\},\\
\mathcal{P}_6&=\{\x_1^*, ~\x_2,~ \x_2,~ \x_2, ~-\x_2,~-\x_2,~-\x_2,~\x_1^*\},\\
\mathcal{P}_7&=\{\x_1, ~\x_2^*,~ \x_2^*,~ \x_2^*, ~-\x_2^*,~-\x_2^*,~-\x_2^*,~-\x_1\},
\end{align*}
and the superscript $*$ indicates inactive nodes. The patterns $\mathcal{P}_1-\mathcal{P}_4$ are illustrated in Fig.~\ref{fig:time_series}. Therefore, the intermediate patterns encountered during the transition from the symmetry-broken state (the initial pattern) to the complete amplitude or oscillation death states are governed by the intercluster weights in the initial pattern $\mathcal{P}_3$. 

If ${\tilde\sigma}(\mathcal{C}_i^{a(o)})$ denotes the RHS in Eqs.~\eqref{eq:amp_death_crit_coup_2} and \eqref{eq:osc_death_crit_coup}, the critical $\sigma$ at which an $i$th cluster $\mathcal{C}_i^{a(o)}$ makes a transition to the amplitude(oscillation) death state, indicated by the superscripts $a(o)$, the existence regime of different patterns for $W_1>W_2$ is as follows 
\begin{alignat*}{2}
&{\tilde{\sigma}}({\mathcal{C}_1^a}) && \le \sigma(\mathcal{P}_2) < {\tilde{\sigma}}({\mathcal{C}_1^o}),\\
&{\tilde{\sigma}}({\mathcal{C}_2^a}) && \le \sigma(\mathcal{P}_3) < {\tilde{\sigma}}({\mathcal{C}_1^o}),\\
&{\tilde{\sigma}}({\mathcal{C}_1^o}) && \le \sigma(\mathcal{P}_4) < {\tilde{\sigma}}({\mathcal{C}_2^o}),
\end{alignat*}
while pattern $\mathcal{P}_1$ exist for all $\sigma$ values. In the same way, starting with a different pattern, we can find the existence of other valid patterns.

Note that none of the patterns discussed above can be identified based on permutation symmetries \cite{kumar2025symmetry}. For general coupling functions for which $\H(\x,\x) \ne 0 \in \mathbb{R}^m$, it has been shown that patterns of coexisting active-inactive clusters arise from permutation symmetries \cite{kumar2025symmetry}. In the network shown in Fig.~\ref{fig:time_series}, the permutation symmetries partition the network into invariant clusters $\{1,6,8\}$, $\{2,3\}$, $\{4\}$ and $\{5,7\}$; however, the symmetry breaking of these clusters does not yield any valid pattern of coexisting active-inactive clusters or antisynchronized active clusters. Therefore, our work shows that dynamically valid patterns of coexisting active-inactive clusters or antisynchronized clusters exist in networks even when the underlying permutation symmetries are insufficient to generate them. 
\begin{figure}
 \includegraphics[width=\linewidth]{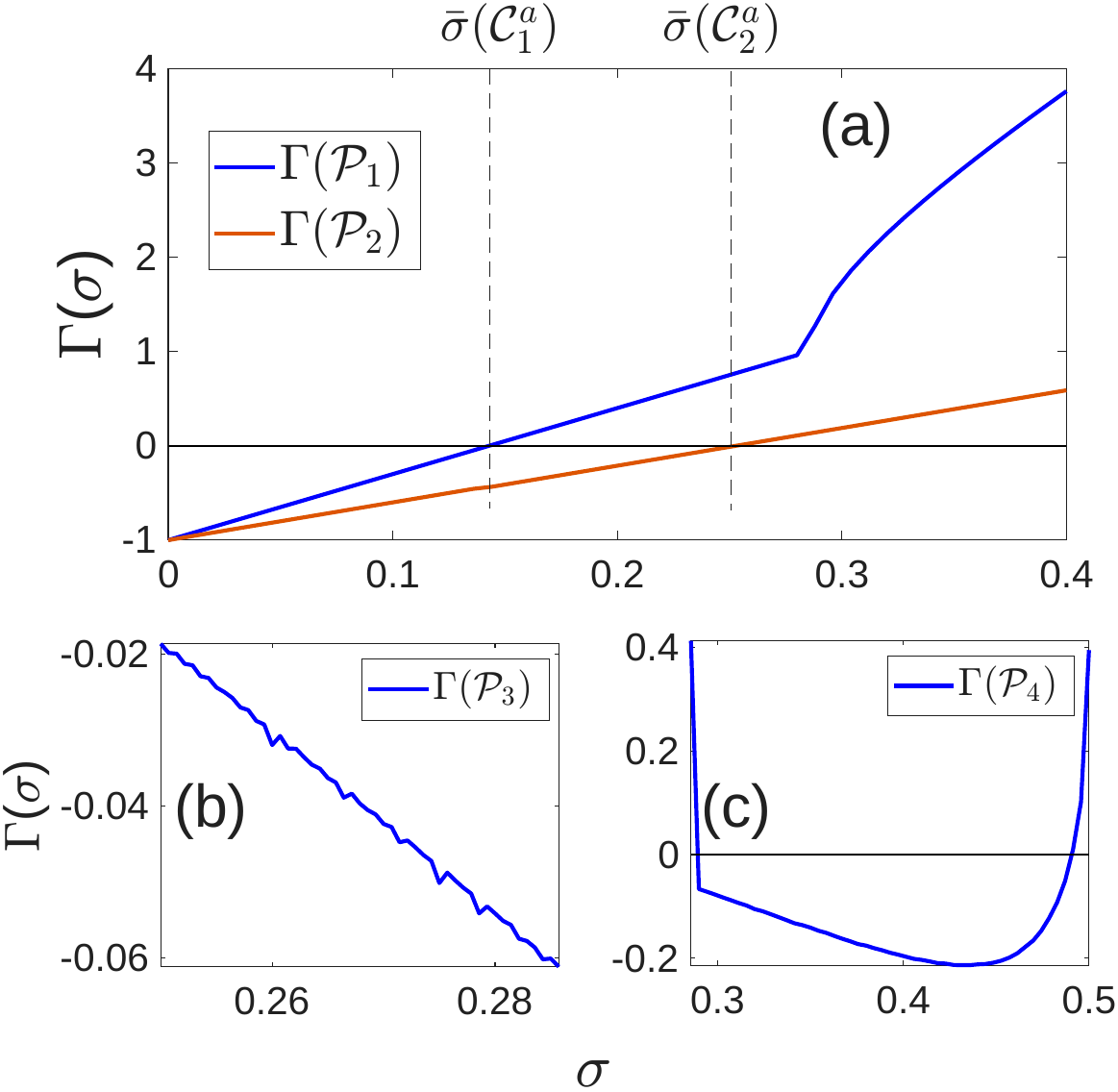}
 \caption{Stability analysis of activity patterns observed in Fig.~\ref{fig:time_series}. A negative $\Gamma(\sigma)$ value for an invariant pattern shows its stability.}
 \label{fig:lyap_exp}
 \end{figure}
\subsection{Stability}\label{sec:stablity_sl}

Finally, we compute the Lyapunov exponent $\Gamma(\sigma)$ for patterns $\mathcal{P}_1-\mathcal{P}_4$ using Eqs.~\eqref{eq:quotient_dyn} and \eqref{eq:perturb_dyn_xi_coordinates} and the resulting graphs are shown in Fig.~\ref{fig:lyap_exp}. Combining the $\sigma$ intervals over which each pattern exists and $\Gamma(\sigma)$ values are negative from Fig.~\ref{fig:lyap_exp}, the stability ranges of patterns $\mathcal{P}_1-\mathcal{P}_4$ is as follows 
\[
\begin{alignedat}{5}
&{} 
&& {} 
&& \sigma(\mathcal{P}_1) 
&& < \,
&& \tilde{\sigma}(\mathcal{C}_1^a),\\
&\tilde{\sigma}(\mathcal{C}_1^a) 
&& \le \,
&& \sigma(\mathcal{P}_2) 
&& < \,
&& \tilde{\sigma}(\mathcal{C}_2^a),\\
&\tilde{\sigma}(\mathcal{C}_2^a) 
&& \le \,
&& \sigma(\mathcal{P}_3) 
&& < \,
&& \tilde{\sigma}(\mathcal{C}_1^o),\\
&\tilde{\sigma}(\mathcal{C}_1^o) 
&& \lessapprox \,
&& \sigma(\mathcal{P}_4) 
&& \lessapprox \,
&& \tilde{\sigma}(\mathcal{C}_2^o).
\end{alignedat}
\]
$\Gamma(\sigma)$ values for patterns $\mathcal{P}_3$ and $\mathcal{P}_4$ are plotted separately (Figs.~\ref{fig:lyap_exp}(b) and \ref{fig:lyap_exp}(c)) only in the $\sigma$ range in which these patterns exist because the corresponding $\Gamma(\sigma)$ values remain negative throughout the $\sigma$ range, except for two bumps in $\Gamma(\mathcal{P}_4)$, Fig.~\ref{fig:lyap_exp}(c).

\section{Conclusion}\label{sec:conclusions}

We have demonstrated dynamically valid activity patterns of coexisting active and inactive clusters of synchronized nodes in networks of identical coupled oscillators, even when the underlying network structure lacks permutation symmetries. The requirement of permutation symmetries for the emergence of synchrony clusters is a strong and often unrealistic constraint for real-world networks. Therefore, establishing the existence of such patterns in networks lacking symmetries is of fundamental importance. We have shown that a sufficient condition for a dynamical system to exhibit coexisting active and inactive clusters is that its intrinsic and coupling dynamics are odd functions in phase space. Starting with the complete amplitude death state, we have systematically identified all possible patterns in amplitude or oscillation death states from symmetry breaking of synchronized clusters in amplitude death states. We have shown that, besides external equitable partitions (EEPs) based clusters, purely dynamics-induced clusters in the amplitude death state can also arise. Some of the invariant patterns may be transient, so we have presented a method to identify the coupling range in which these patterns exist for all times. Our work shows that transition pathways, intermediate patterns through which a network traverses as coupling changes, are controlled by intercluster weights. Numerical simulations have been presented for the paradigmatic Stuart-Landau oscillators connected through Laplacian coupling. Finally, by combining the synchronization manifold with Laplacian eigenvectors, we have identified transversal perturbations for these patterns, and a stability analysis has been presented.

In summary, our work demonstrates that networks without permutation symmetries can nonetheless support stable coexistence of active and inactive clusters of synchronized nodes, significantly broadening the class of networks in which such complex collective dynamics can arise.

\begin{acknowledgments}
A.K. acknowledges the financial support from IISER Thiruvananthapuram. VKC acknowledges the  DST,  New Delhi, for computational facilities under the DST-FIST program (Grant No. SR/FST/PS-1/2020/135) to the Department of Physics.  DVS  is supported by the ANRF Project under Grant No. ANRF/ARG/2025/008542/PS.\\
\end{acknowledgments}

\appendix

\section{Invariance of $\mathcal{S}$ and $\mathcal{S}_{\perp}$}\label{sec:app_invariant_decomp}

\begin{proposition}(Invariant decomposition induced by an EEP)
Let $L \in \mathbb{R}^{N \times N}$ be a symmetric network Laplacian admitting an EEP with $k$ clusters, and let $\mathcal{S} \subseteq R^{Nm}$ denote the corresponding synchronization subspace. Then:
\begin{enumerate}

\item The synchronization subspace $\mathcal{S}$ defined as 
\begin{align*}\label{}
\mathcal{S}=\{\x \in \mathbb{R}^{Nm}: \x_i=\s_l \in \mathbb{R}^m,~\forall ~i \in \mathcal{C}_l, ~\forall ~l \in [k]\}
\end{align*}
is invariant under operator $\mathcal{L}$, i.e., $\mathcal{L}\mathcal{S} \subseteq \mathcal{S}$, where $\mathcal{L}$ is defined from $\dot {\deltax}=\mathcal{L}\deltax$ (Eq.~\eqref{eq:perturb_dyn}).

\item The orthogonal complement $\mathcal{S}_{\perp}$ defined as
\begin{align*}\label{}
\mathcal{S}_{\perp}=\{\x \in \mathbb{R}^{Nm}: \sum_{j ~\in~\mathcal{C}_l}\x_j=0 \in \mathbb{R}^m,~\forall ~l \in [k]\}
\end{align*}
    is also invariant under $\mathcal{L}$, i.e., $\mathcal{L} \mathcal{S}_{\perp} \subseteq \mathcal{S}_{\perp}$.
\end{enumerate}
\end{proposition}

\paragraph*{Proof.}
First, we prove that the synchronization subspace $\mathcal{S}$ is invariant under $\mathcal{L}$.
If we choose $\deltax \in \mathcal{S}$, the first term in $\mathcal{L}$ is invariant because $D\F(\s_l)$ is identical for all nodes in any cluster $\mathcal{C}_l$. Therefore, $(E^l \otimes D\F(\s_l)) ~\deltax \in \mathcal{S} ~\forall ~l$. The second term in $\mathcal{L}$ is also invariant by definition of an EEP, as nodes belonging to the same cluster receive identical total coupling from every other cluster. Hence, for any vector $\deltax \in \mathcal{S}$ which is constant on each cluster, $(LE^l \otimes D\H(\s_l)) ~\deltax \in \mathcal{S} ~\forall ~l$. Therefore,
\begin{equation*}
\mathcal{L} \mathcal{S} \subseteq \mathcal{S}.
\end{equation*}
Next, we prove that the orthogonal complement $\mathcal{S}_{\perp}$ is invariant under $\mathcal{L}$. Note that the property of zero sum over each cluster, $\sum_{j ~\in~\mathcal{C}_l}\x_j=0 ~\forall~l \in [k]$ follows from the orthogonality of $\mathcal{S}_{\perp}$ to $\mathcal{S}$ as $Q^T\mathcal{S}_{\perp}=0$. Let $\deltax\in \mathcal{S}_\perp$, and because $D\F(\s_l)$ is identical for all nodes in a cluster, $(E^l \otimes D\F(\s_l)) ~\deltax \in \mathcal{S}_{\perp} ~\forall ~l$.  Now, only the invariance due to the coupling terms is required. An $i$th row in the column vector $\mathcal{L}\deltax$ is 
\begin{equation*}
({\mathcal L}\,\deltax)_i
=
\sum_{l=1}^k
\sum_{j\in \mathcal{C}_l}
L_{ij}\,
D\H(\s_l)\,\deltax_j .
\end{equation*}
Since $D\H(\s_l)$ is same for all $j \in \mathcal{C}_l$, we can factor it out
\begin{equation*}
({\mathcal L}\,\deltax)_i
=
\sum_{l=1}^k
D\H(\s_l)\sum_{j\in \mathcal{C}_l}
L_{ij}\,\deltax_j .
\end{equation*}
We compute the cluster sum of
${\mathcal L}\,\deltax$ over a fixed cluster $\mathcal{C}_r$:
\begin{align*}
\sum_{i\in C_r} (\mathcal{L}\,\deltax)_i
&=
\sum_{l=1}^k
D\H(\s_l)
\sum_{j\in \mathcal{C}_l}
\left(
\sum_{i\in \mathcal{C}_r} L_{ij}
\right)
\deltax_j .
\end{align*}
Since the partition is an EEP and the Laplacian is symmetric, the inner sum $\sum_{i\in C_r} L_{ij}$ is constant for all $j \in \mathcal{C}_l$, and we call this $c_{rl}$. Therefore,
\begin{equation*}
\sum_{i\in \mathcal{C}_r} ({\mathcal L}\,\deltax)_i
=
\sum_{l=1}^k
D\H(\s_l)\,
c_{rl}
\sum_{j\in \mathcal{C}_l} \deltax_j
= 0,
\end{equation*}
because $\deltax\in \mathcal{S}_\perp$ implies
$\sum_{j\in \mathcal{C}_l} \deltax_j = 0$ for all $l$.

Hence,
\begin{equation*}
{\mathcal L}\mathcal{S}_\perp\subseteq \mathcal{S}_\perp,
\end{equation*}
which completes the proof.

\bibliography{references}

\end{document}